\documentclass{ws-procs9x6} 
\begin{document} 
\title{OVERVIEW OF STRANGENESS NUCLEAR PHYSICS} 
\author{AVRAHAM GAL} 
\address{Racah Institute of Physics, The Hebrew University, Jerusalem 91904, 
ISRAEL} 
\begin{abstract} 
Selected topics in Strangeness Nuclear Physics are reviewed: 
$\Lambda$-hypernuclear spectroscopy and structure, 
multistrangeness, and $\overline K$ mesons in nuclei. 
\end{abstract} 


\bodymatter

\section{Introduction}
\label{sec:intro}

The properties of hypernuclei reflect the nature of the underlying
baryon-baryon interactions and, thus, can provide tests of models 
for the free-space hyperon-nucleon ($YN$) and hyperon-hyperon ($YY$) 
interactions. The Nijmegen group has constructed a number of meson-exchange, 
soft-core models, using SU(3)$_{\rm f}$ symmetry to relate coupling constants 
and form factors \cite{rijken08}. The J\"ulich group, in addition to $YN$ 
meson exchange models \cite{haidenbauer05}, published recently leading-order 
chiral effective-field theory $YN$ and $YY$ potentials \cite{polinder06}. 
Quark models have also been used within the $(3q)-(3q)$ resonating group 
model (RGM), augmented by a few effective meson exchange potentials of scalar 
and pseudoscalar meson nonets directly coupled to quarks\cite{fujiwara07}. 
Finally, we mention recent lattice QCD calculations \cite{beane07,nemura09}. 

On the experimental side, there is a fair amount of data on single-$\Lambda$ 
hypernuclei, including production, structure and decay modes \cite{hashim06}. 
Little is known on strangeness $S$=$-2$ hypernuclei. The missing information 
is vital for extrapolating into strange hadronic matter, for both finite 
systems and in bulk, and into neutron stars \cite{jsb08}. 
Therefore, following a brief review of single-$\Lambda$ hypernuclei in 
Sect.~\ref{sec:lambda}, and even a briefer review of $\Sigma$-hyperon nuclear 
interactions in Sect.~\ref{sec:sigma}, I discuss in Sect.~\ref{sec:shm} 
the onset of hyperon nuclear binding, through which the strength of $YN$ and 
$YY$ interactions may be determined; and aspects of $\overline K$ nuclear 
interactions in Sect.~\ref{sec:Kbar}, highlighting the issue of kaon 
condensation. As for the list of references at the very end, many of the 
recent ones require adding `and references cited therein'.  

\section{$\Lambda$ Hypernuclei} 
\label{sec:lambda} 

\begin{figure}[htp]
\centerline{
\includegraphics[width=6.0cm]{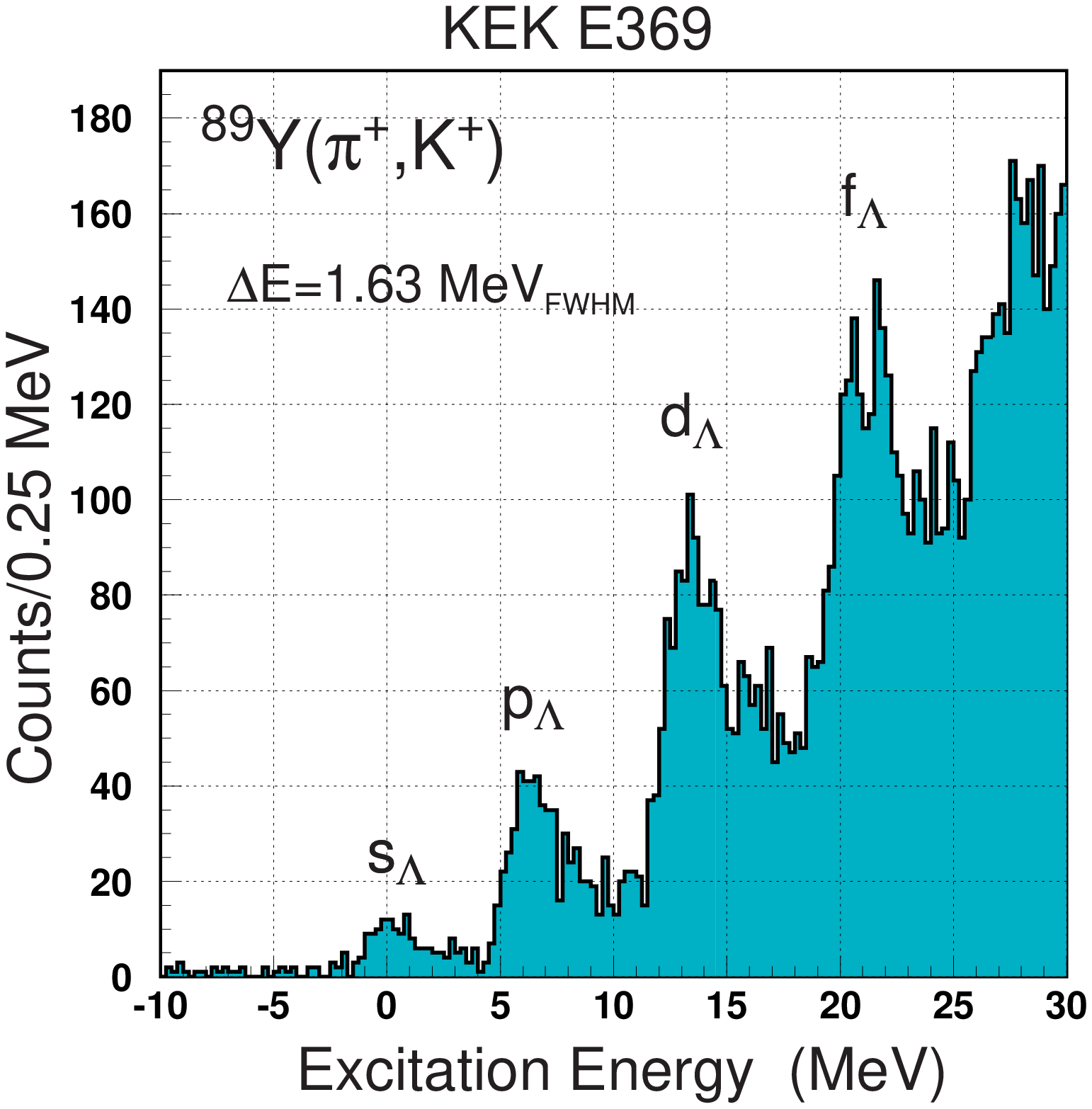}
\includegraphics[width=6.0cm]{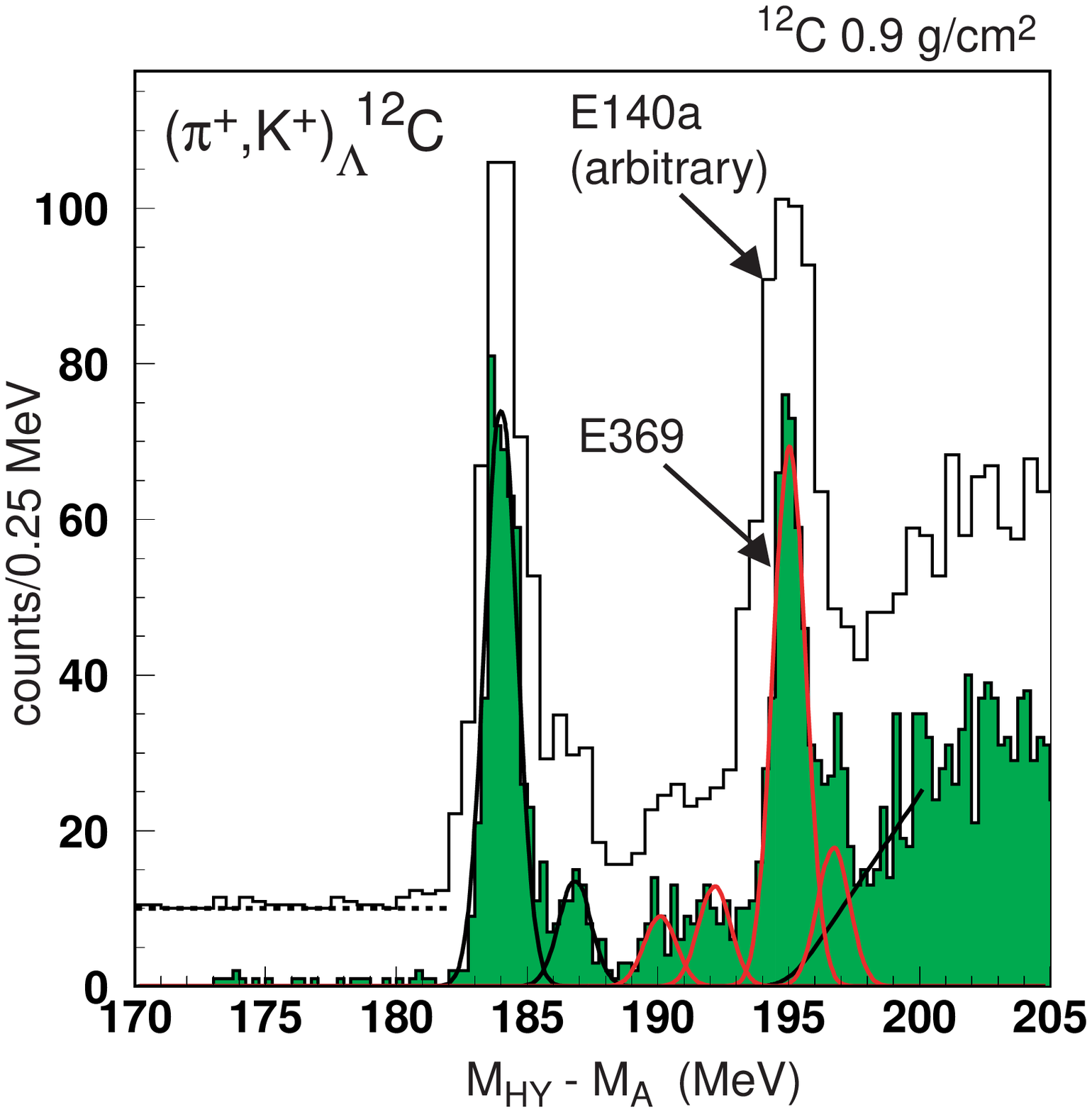}} 
\caption{($\pi^+,K^+$) spectra of $\Lambda$ hypernuclei, from 
Ref.~\cite{hotchi01}.}  
\label{fig:s.p.}
\end{figure} 

To test $YN$ models against the considerable body of information on $\Lambda$ 
hypernuclei, effective interactions for use in limited spaces of shell-model 
orbits must be evaluated. The $\Lambda$ well depth resulting from soft-core 
Nijmegen nuclear-matter $G$-matrices \cite{rijken08,yam08} can be brought to 
a reasonable agreement with the empirical value 28 MeV deduced in fitting 
binding energies of $\Lambda$ single-particle (s.p.) states \cite{mdg88}. 
However, the partial-wave contributions, in particular the spin dependence 
of the central interaction, vary widely in different models, and the 
$\Lambda$-nuclear spin-orbit splitting does not come sufficiently small in 
any of the published models{\footnote{Nevertheless, it was suggested recently 
that missed $\Lambda\to \Sigma\to \Lambda$ iterated one-pion exchange 
contributions cancel out the short-range $\sigma + \omega$ mean-field 
normal contributions to the $\Lambda$ nuclear spin-orbit potential. 
\cite{kaiser08}}}. Figure~\ref{fig:s.p.} shows examples of $\Lambda$ s.p. 
peak structures in $_{~\Lambda}^{89}{\rm Y}$ and in $_{~\Lambda}^{12}{\rm C}$. 
Although the splitting of the $f_{\Lambda}$ orbit in $_{~\Lambda}^{89}{\rm Y}$ 
may suggest a spin-orbit splitting of 1.7 MeV, a more careful shell-model 
analysis shows that it is consistent with a $\Lambda$ spin-orbit splitting 
of merely 0.2 MeV, with most of the observed splitting due to mixing of 
different $\Lambda N^{-1}$ particle-hole excitations \cite{motoba08}. 
Interesting hypernuclear structure is also revealed between major $\Lambda$ 
s.p. states in $_{~\Lambda}^{12}{\rm C}$. This has not been studied yet with 
sufficient resolution in medium-weight and heavy hypernuclei, but data already 
exist from JLab on $^{12}$C and other targets, with sub-MeV resolution, 
as shown in this Symposium by Garibaldi and Tang. Furthermore, even with the 
coarser resolution of the $(\pi^+,K^+)$ data shown in Fig.~\ref{fig:s.p.}, 
most of the $_{~\Lambda}^{12}{\rm C}$ levels between the (left) $1s_{\Lambda}$ 
peak and the (right) $1p_{\Lambda}$ peak are particle-stable and could be 
studied by looking for their electromagnetic cascade deexcitation to the 
ground state. 

\begin{figure}
\centerline{
\includegraphics[width=4.4in]{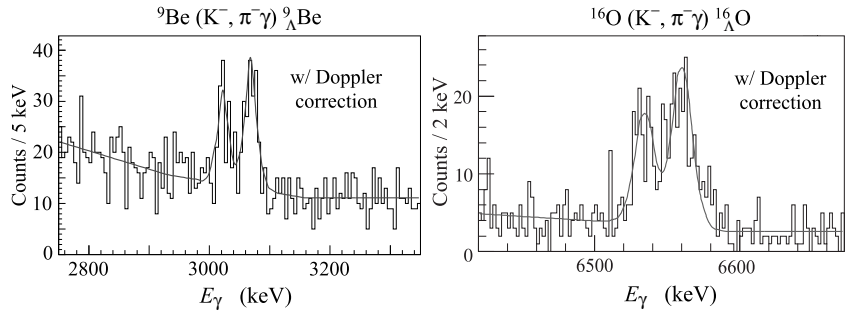}}
\caption{$\gamma$-ray spectra of $\Lambda$ hypernuclei from BNL-E930, see 
Tamura's review~\cite{tamura08}. The observed twin peaks (in order left to 
right) result from the ${\frac{5}{2}}^+$ and ${\frac{3}{2}}^+$ levels in 
$_{\Lambda}^9{\rm Be}$ separated by 43 keV, deexciting to the ground state, 
and from deexcitation of a $1^{-\star}$ level in $_{~\Lambda}^{16}{\rm O}$ 
to the ground-state doublet $0^-$ and $1^-$ levels separated by 26 keV. } 
\label{fig:9be16o}
\end{figure}

A systematic program of $\gamma$-ray measurements has been carried out 
for light $\Lambda$ hypernuclei at BNL and KEK \cite{tamura08} following 
a proposal made long time ago by Dalitz and Gal \cite{dg78}, in order to 
study the spin dependence of the {\it effective} $\Lambda N$ interaction 
in the nuclear $p$ shell,  
\begin{equation} 
V_{\Lambda N} = \bar{V}+\Delta\,{\vec s}_N\cdot {\vec s}_\Lambda+S_\Lambda\, 
{\vec l}_N\cdot {\vec s}_\Lambda + S_N\,{\vec l}_N\cdot {\vec s}_N + T\,S_{12} 
\,, 
\label{eq:spin} 
\end{equation} 
specified in terms of four radial matrix elements: $\Delta$ for spin-spin, 
$S_\Lambda$ and $S_N$ for spin-orbit, $T$ for the tensor interaction. 
The most completely studied hypernucleus todate is $_{\Lambda}^7{\rm Li}$ with 
five observed $\gamma$-ray transitions, allowing a good determination of these 
parameters in the beginning of the $p$ shell \cite{millener08}: 
\begin{equation} 
A=7,9: \,\,\,\,\,\, 
\Delta=430,\,\, S_\Lambda=-15,\,\, S_N=-390,\,\, T=30 \,\,\,({\rm keV}). 
\label{eq:7-9} 
\end{equation} 
The dominant contributions to $_{\Lambda}^7{\rm Li}$ level spacings 
are due to $\Delta$ for $1s_\Lambda$ inter-doublet spacings, and $S_N$ for 
intra-doublet spacings (c.f. Table~\ref{tab:doublet}). 

A remarkable experimental observation of minute doublet spin splittings 
in $_{\Lambda}^9{\rm Be}$ and in $_{~\Lambda}^{16}{\rm O}$ is shown in 
Fig.~\ref{fig:9be16o}. The contributions of the various spin-dependent 
components of the effective $\Lambda N$ interaction to these doublet 
splittings are given in Table~\ref{tab:doublet} using Eq.~(\ref{eq:7-9}) 
for $_{\Lambda}^9{\rm Be}$ and a somewhat revised parameter set for heavier 
hypernuclei, in the $p_{\frac{1}{2}}$ subshell, which exhibit greater 
sensitivity to the tensor interaction \cite{millener08}: 
\begin{equation} 
A=15,16: \,\,\,\,\,\,
\Delta=315,\,\, S_\Lambda=-15,\,\, S_N=-350,\,\, T=23.2 \,\,\,({\rm keV}). 
\label{eq:11-15} 
\end{equation} 
Listed also are calculated $\Lambda\Sigma$ mixing contributions, as detailed 
in Ref.~\cite{millener08}. Very small core polarization contributions bounded 
by 10 keV are not listed. In $_{\Lambda}^9{\rm Be}$, since both $\Delta$ and 
$T$ are well controlled by data from other systems, it is fair to state that 
the observed $43\pm 5$ keV doublet splitting provides a stringent measure of 
the smallness of the $\Lambda$ spin-orbit term in $\Lambda$ hypernuclei, 
consistently with the small $p_{\frac{1}{2}} - p_{\frac{3}{2}}$ $\Lambda$ 
spin-orbit splitting associated with the $\Delta E = 152\pm 54({\rm stat.})
\pm 36({\rm syst.})$ keV splitting observed in $_{~\Lambda}^{13}{\rm C}$ 
\cite{ajimura01}. 

\begin{table}
\tbl{Contributions calculated by Millener \cite{millener08} of $\Lambda\Sigma$ 
mixing and $\Lambda N$ spin-dependent interaction terms, Eq.~(\ref{eq:spin}), 
to doublet splittings in $_{\Lambda}^7{\rm Li}$ and $_{\Lambda}^9{\rm Be}$ 
using Eq.~(\ref{eq:7-9}), and in $_{~\Lambda}^{15}{\rm N}$ and 
$_{~\Lambda}^{16}{\rm O}$ using Eq.~(\ref{eq:11-15}), are compared with 
experiment \cite{tamura08} (in keV). } 
{\begin{tabular}{@{}lccccccccc@{}}\toprule 
$_{\Lambda}^{Z}A$ & $J_{\rm upper}$ & $J_{\rm lower}$ & $\Lambda\Sigma$ & 
$\Delta$ & $S_\Lambda$ & $S_{\rm N}$ & T & $\Delta E_{\rm calc.}$ &
$\Delta E_{\rm exp.}$ \\\colrule 
$_{\Lambda}^7{\rm Li}$ & ${\frac{3}{2}}^+$ & ${\frac{1}{2}}^+$ & $72$ & $628$ 
& $-1$ & $-4$ & $-9$ & $693$ & $691.7\pm 1.2$ \\ 
$_{\Lambda}^9{\rm Be}$ & ${\frac{3}{2}}^+$ & ${\frac{5}{2}}^+$ & $-8$ & $-14$ 
& $37$ & $0$ & $28$ & $44$ & $43\pm 5$ \\ 
$_{~\Lambda}^{15}{\rm N}$ & ${\frac{1}{2}}^+$ & ${\frac{3}{2}}^+$ & $42$ & 
$232$ & $34$ & $-8$ & $-208$ & $92$ & \\  
$_{~\Lambda}^{16}{\rm O}$ & $1^-$ & $0^-$ & $-29$ & $-117$ & $-21$ & $1$ & 
$183$ & $27$ & $26.4 \pm 1.7$ \\ \botrule
\end{tabular}}
\label{tab:doublet} 
\end{table} 

The spin dependence of the $\Lambda N$ interaction may also be studied 
by observing pionic weak-decay spectra, as reported in this Symposium 
by Botta for the FINUDA Collaboration \cite{finuda09}. In particular, 
the $_{~\Lambda}^{15}{\rm N} \to \pi^-+{^{15}{\rm O}}$ measured spectrum 
suggests a spin-parity assignment 
$J^{\pi}(_{~\Lambda}^{15}{\rm N}_{\rm g.s.})={\frac{3}{2}}^+$, consistently 
with the positive value predicted by Millener \cite{millener08} for the 
ground-state doublet splitting $E({\frac{1}{2}}^+)-E({\frac{3}{2}}^+)$ 
listed in Table~\ref{tab:doublet}.

\section{$\Sigma$ hyperons} 
\label{sec:sigma} 

\begin{figure}[htp]
\centerline{
\includegraphics[width=5.8cm,height=6.5cm]{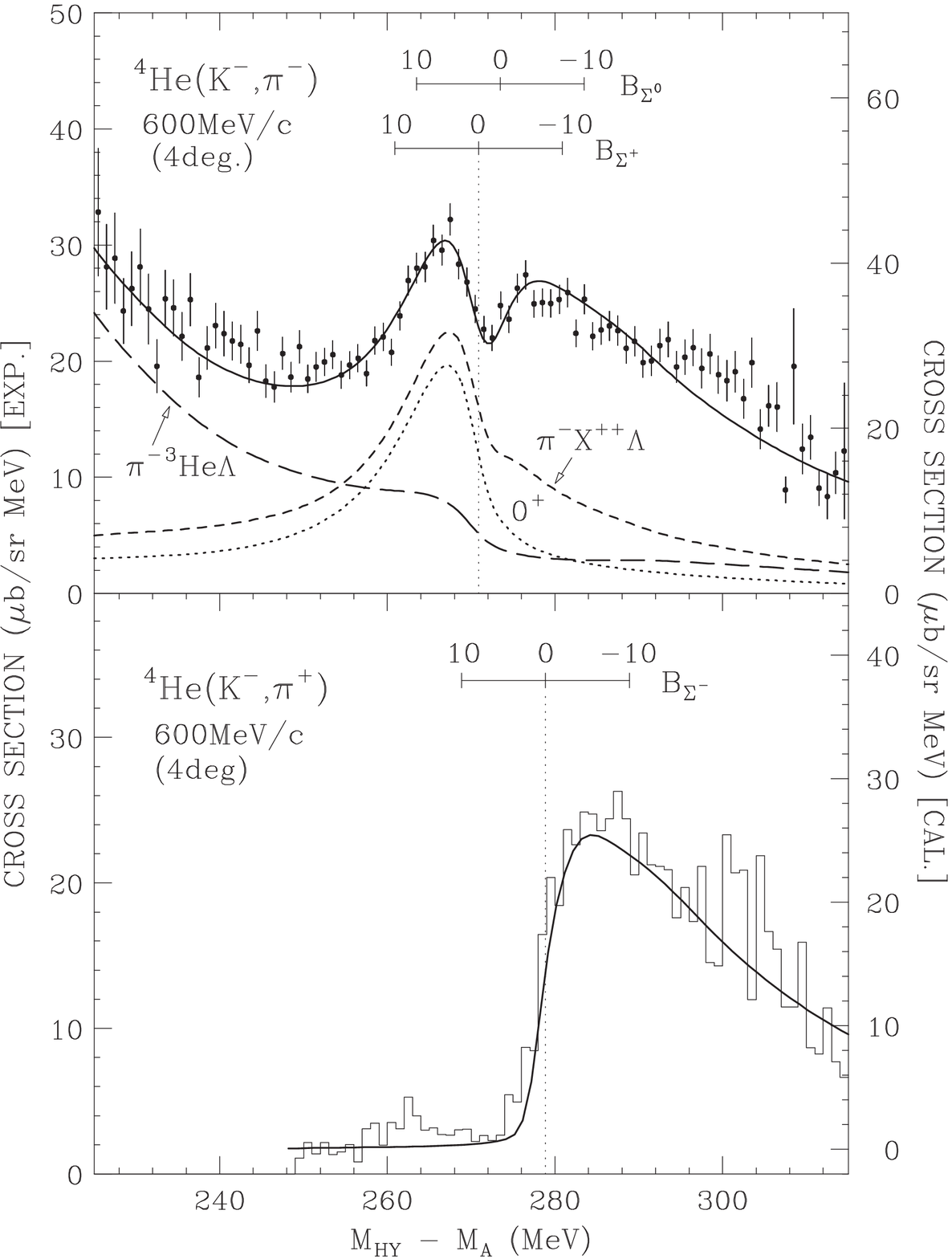}
\includegraphics[width=5.3cm,height=6.5cm]{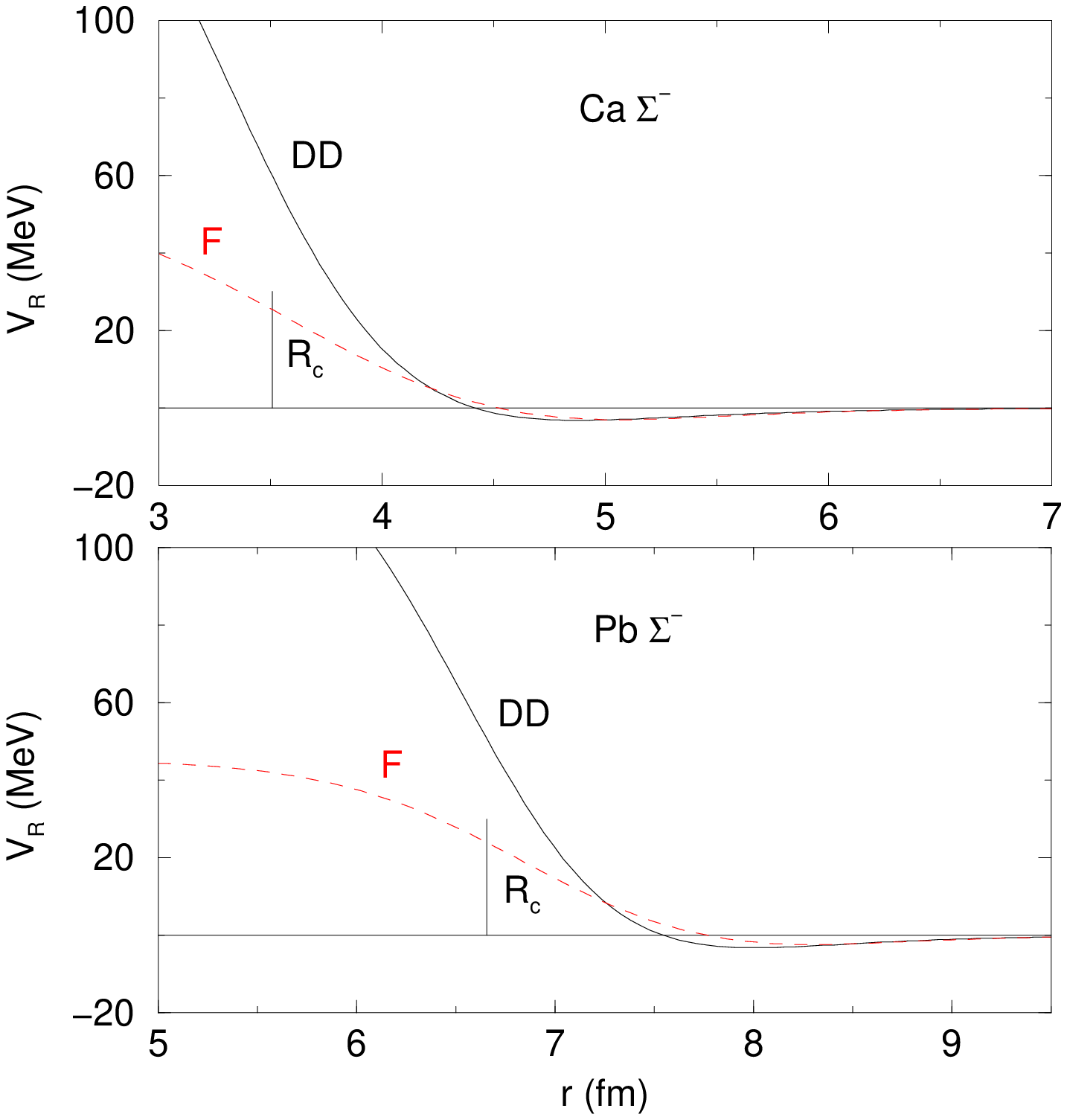}}
\caption{Left: $^4{\rm He}(K^-,\pi^{\pm})$ spectra, as measured \cite{nagae98} 
and as calculated by Harada \cite{harada98}, providing evidence for 
a $^4_\Sigma$He $I=1/2$ quasibound state in the $\pi^-$ channel, 
with binding energy $B_{\Sigma^+} =4.4 \pm 0.3 \pm 1$ MeV and width 
$\Gamma=7.0 \pm 0.7^{+1.2}_{-0.0}$ MeV. Right: Re~$V_{\rm opt}^{\Sigma}$ 
fitted to {\it all} $\Sigma^-$ atomic data, for two density-dependent 
potential models \cite{fg07}. The half-density nuclear charge radius 
$R_c$ is indicated.} 
\label{fig:sigma}
\end{figure} 

A vast body of reported $(K^-,\pi^{\pm})$ and $(\pi^-,K^+)$ spectra indicate 
a repulsive $\Sigma$ nuclear potential, with a substantial isospin dependence 
\cite{dmg89} which for very light nuclei may conspire in selected 
configurations to produce $\Sigma$ hypernuclear quasibound states, as shown on 
the left-hand side (l.h.s.) of Fig.~\ref{fig:sigma} for $^4_\Sigma$He.\footnote
{The discovery of $^4_\Sigma$He, in $K^-$ capture at rest, is due to Hayano 
{\it et al.} \cite{hayano89}} These data, including recent $(\pi^-,K^+)$ 
spectra \cite{noumi02} and related DWIA analyses \cite{kohno04}, suggest that 
$\Sigma$ hyperons do not bind in heavier nuclei. 

A repulsive component of the $\Sigma$ nuclear potential is also revealed in 
analyses of strong-interaction level shifts and widths in $\Sigma^-$ atoms, 
as shown on the right-hand side (r.h.s.) of Fig.~\ref{fig:sigma}. In fact, 
Re~$V_{\rm opt}^{\Sigma}$ is attractive at low densities outside the nucleus, 
as enforced by the observed `attractive' $\Sigma^-$ atomic level shifts, 
changing into repulsion well outside of the nuclear radius. The precise 
magnitude and shape of the repulsive component within the nucleus, however, 
are model dependent \cite{fg07}. The slightly prefered potential F yields 
Re~$V_{\rm opt}^{\Sigma}(\rho_0)\sim 40-50$~MeV, roughly consistent with 
Refs.~\cite{kohno04}. This bears interesting consequences for the balance of 
strangeness in the inner crust of neutron stars, primarily by delaying the 
appearance of $\Sigma^-$ hyperons to higher densities, as shown on the l.h.s. 
of Fig.~\ref{fig:sbg} in Sect.~\ref{sec:shm}.   

\begin{table} 
\tbl{Isoscalar and isovector hyperon potentials, Eq.~(\ref{eq:lane}) in MeV, 
calculated for Nijmegen soft-core potential models \cite{rijken08,rijken06}, 
denoted by year and version, at nuclear-matter density 
($k_F=1.35~{\rm fm}^{-1}$). The ESC06 results are preliminary. 
The ESC$^{\star}$ models assume specifically repulsive medium modifications 
affecting weakly the isovector potentials. Excluded are Im~$V^{\Sigma}$ due 
to $\Sigma N\to \Lambda N$ and Im~$V^{\Xi}$ due to $\Xi N\to \Lambda\Lambda$.} 
{\begin{tabular}{@{}lccccccccc@{}}\toprule  
 & 97f & 04a & 04a$^{\star}$ & 04d & 04d$^{\star}$ & 06d & 06d$^{\star}$ 
& phenom. & Ref. \\\colrule
$V_0^{\Lambda}$ & $-31.7$ & $-38.5$ & $-30.6$ & $-44.1$ & $-37.2$ & $-44.5$ & 
$-37.5$ & $-28$ & \cite{mdg88} \\ 
$V_0^{\Sigma}$ & $-13.9$ & $-36.5$ & $-27.9$ & $-26.0$ & $-16.6$ & $-1.2$ & 
$+8.2$ & $10-50$ & \cite{fg07,kohno04} \\ 
$V_1^{\Sigma}$ & $-30.4$ & $+21.6$ & & $+30.4$ & & $+52.6$ & $+55.2$ & 
$\approx +80$ & \cite{dgm84} \\ 
$V_0^{\Xi}$ & & $+15.1$ & & $-18.7$ & $-12.1$ & & & 
$\approx -14$ & \cite{fukuda98} \\ 
$V_1^{\Xi}$ & & $+32.5$ & & $+50.9$ & $+51.5$ & & & & \\\botrule
\end{tabular}} 
\label{tab:sig} 
\end{table} 

The $G$-matrices constructed from Nijmegen soft-core potential models 
generally do not produce $\Sigma$ repulsion in symmetric nuclear matter, 
as demonstrated in Table~\ref{tab:sig} using the parametrization 
\begin{equation} 
\label{eq:lane} 
V^Y = V_0^Y + \frac{1}{A}~V_1^Y~{\bf T}_A{\cdot}{\bf t}_Y ~. 
\end{equation} 
In contrast to the published Nijmegen soft-core attractive potentials, SU(6) 
quark-model RGM calculations \cite{fujiwara07} in which a strong Pauli 
repulsion appears in the $I=3/2,~{^3S_1}-{^3D_1}~\Sigma N$ channel 
give repulsion, and so does an SU(3) chiral perturbation calculation 
\cite{kaiser05} which yields repulsion of order 60 MeV. 
Phenomenologically $V_0^{\Sigma} > 0$ and $V_1^{\Sigma} > 0$, as listed in 
the table, and both components of $V^{\Sigma}$ give repulsion in nuclei. 
However, given a nuclear core with $(N-Z)<0$ and owing to the small 
value of $A$ in $^4_\Sigma$He, the isovector term provides substantial 
attraction towards binding this exceptional hypernucleus $^4_\Sigma$He, 
while the isoscalar repulsion reduces the quasibound level width 
(c.f. Fig.~\ref{fig:sigma}). 

\section{Strangeness binding onset and Strange Hadronic Matter} 
\label{sec:shm} 

\begin{figure}
\centerline{
\includegraphics[width=4.6in]{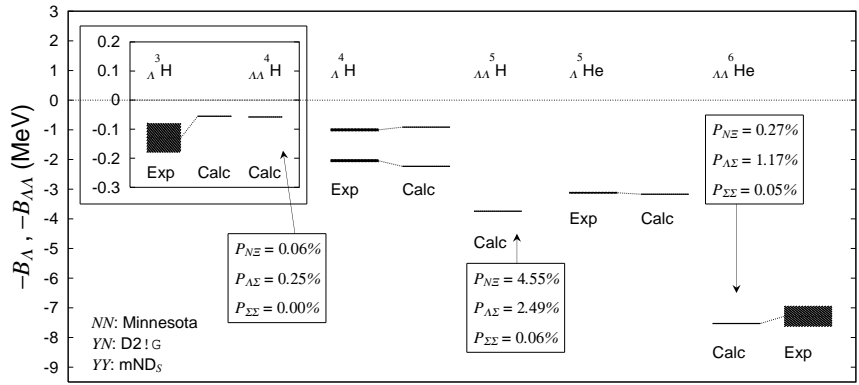}} 
\caption{$\Lambda$ and $\Lambda\Lambda$ separation energies in $s$-shell 
hypernuclei, calculated in Ref.~\cite{nemura05}}   
\label{fig:nemura}
\end{figure} 

Complete few-body calculations of the $s$-shell hypernuclei, for systems of 
nucleons and $\Lambda$ hyperons, with full account of coupled-channel effects 
due to the primary $\Lambda N - \Sigma N$ and $\Lambda \Lambda - \Xi N$ 
mixings, were reported by Nemura {\it et al.} \cite{nemura05} using stochastic 
variational methods and phenomenological potentials based partly on meson 
exchange models. The calculated spectra are shown in Fig.~\ref{fig:nemura}. 
In addition to the established $_{\Lambda}^3{\rm H}$, 
$_{\Lambda}^4{\rm H}-{_{\Lambda}^4{\rm He}}$ and $_{\Lambda}^5{\rm He}$ 
single-$\Lambda$ hypernuclei, $_{\Lambda\Lambda}^{~~4}{\rm H}$ and 
$_{\Lambda\Lambda}^{~~5}{\rm H}$ - $_{\Lambda\Lambda}^{~~5}{\rm He}$ 
bound states were predicted by fitting to 
$\Delta B_{\Lambda\Lambda}(_{\Lambda\Lambda}^{~~6}{\rm He}) \approx 1$~MeV 
for the only $\Lambda\Lambda$ hypernucleus uniquely assigned by experiment 
\cite{nagara01}. We note that $_{\Lambda\Lambda}^{~~4}{\rm H}$ is 
particle-stable in the calculation of Fig.~\ref{fig:nemura} only by a minute 
2 keV; given the uncertainties in the input and in the calculations, 
$_{\Lambda\Lambda}^{~~4}{\rm H}$ could still prove unbound \cite{filikhin02}. 
Moreover, the experimental evidence \cite{ahn01} for 
$_{\Lambda\Lambda}^{~~4}{\rm H}$ has been challenged recently \cite{hunger07}. 
In contrast, the particle stability of $_{\Lambda\Lambda}^{~~5}{\rm H}$ and 
$_{\Lambda\Lambda}^{~~5}{\rm He}$, which have not yet been discovered, appears 
theoretically robust \cite{fg02}.

Very little is established experimentally on the interaction of $\Xi$ hyperons 
with nuclei. Inclusive $(K^-,K^+)$ spectra \cite{fukuda98} on $^{12}$C yield 
a somewhat shallow attractive potential, $V^{\Xi} \approx -14$~MeV, by fitting 
near the $\Xi^-$ hypernuclear threshold. Of the Nijmegen soft-core potentials 
listed in Table~\ref{tab:sig}, ESC04d$^{\star}$ is the closest one to 
reproducing the phenomenological potential depth and it gives rise, 
selectively -- owing to its strong spin and isospin dependence, to quasibound 
$\Xi$ states in several light nuclear targets, beginning with $^7$Li 
\cite{hiyama08}. In this model, $_{\Xi^0}^{~5}{\rm He}$ is unbound. 
For a nuclear-matter width $\Gamma_{\Xi}=12.7$~MeV calculated in model 
ESC04d$^{\star}$, it may not be straightforward to resolve the rich 
spectroscopy predicted for these light nuclear targets. A `day-1' experiment 
at J-PARC on a $^{12}$C target is scheduled soon \cite{takahashi08}. 

\begin{figure}[htp]
\centerline{
\includegraphics[width=5.5cm,height=5.5cm]{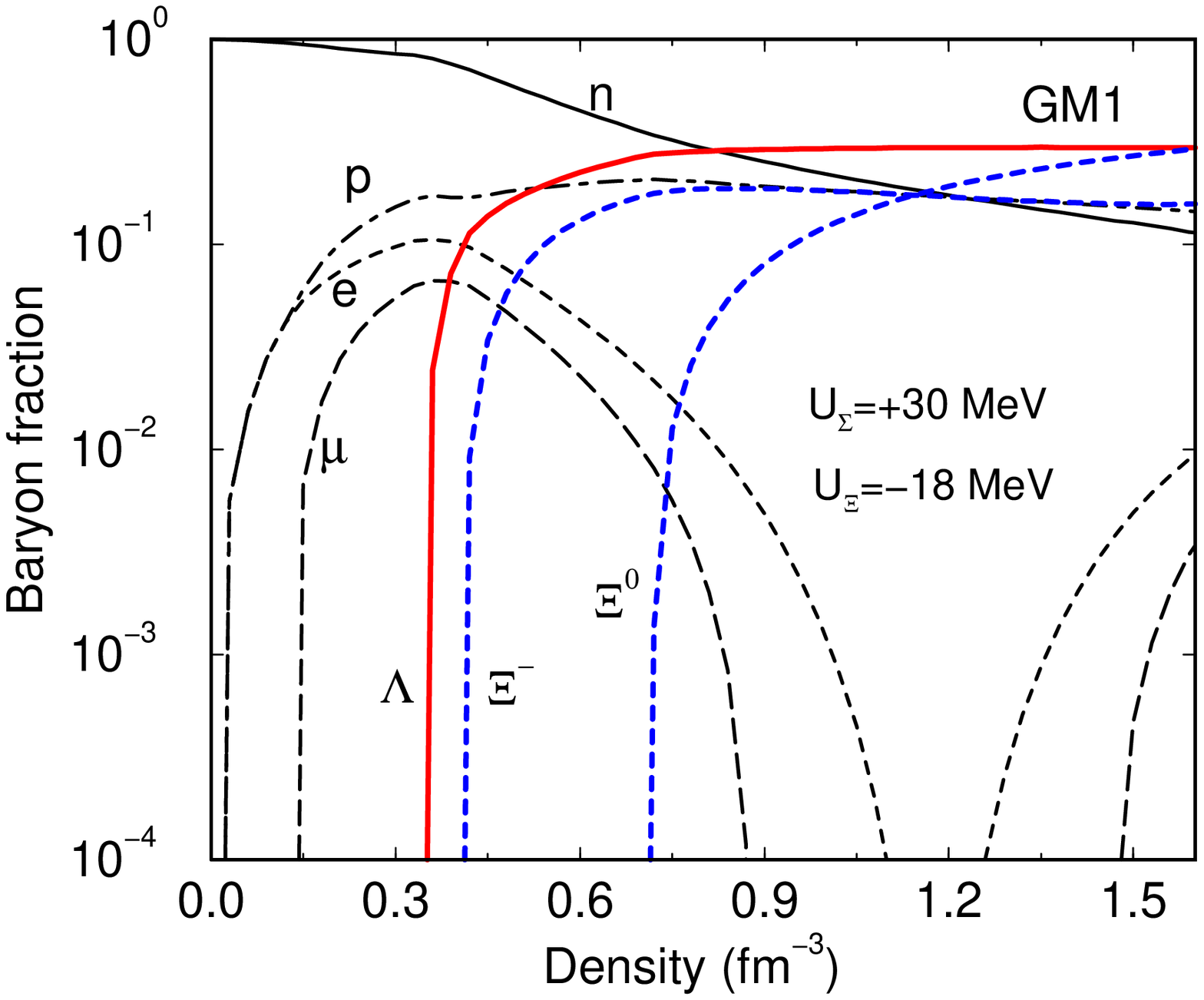} 
\includegraphics[width=6.0cm,height=6.0cm]{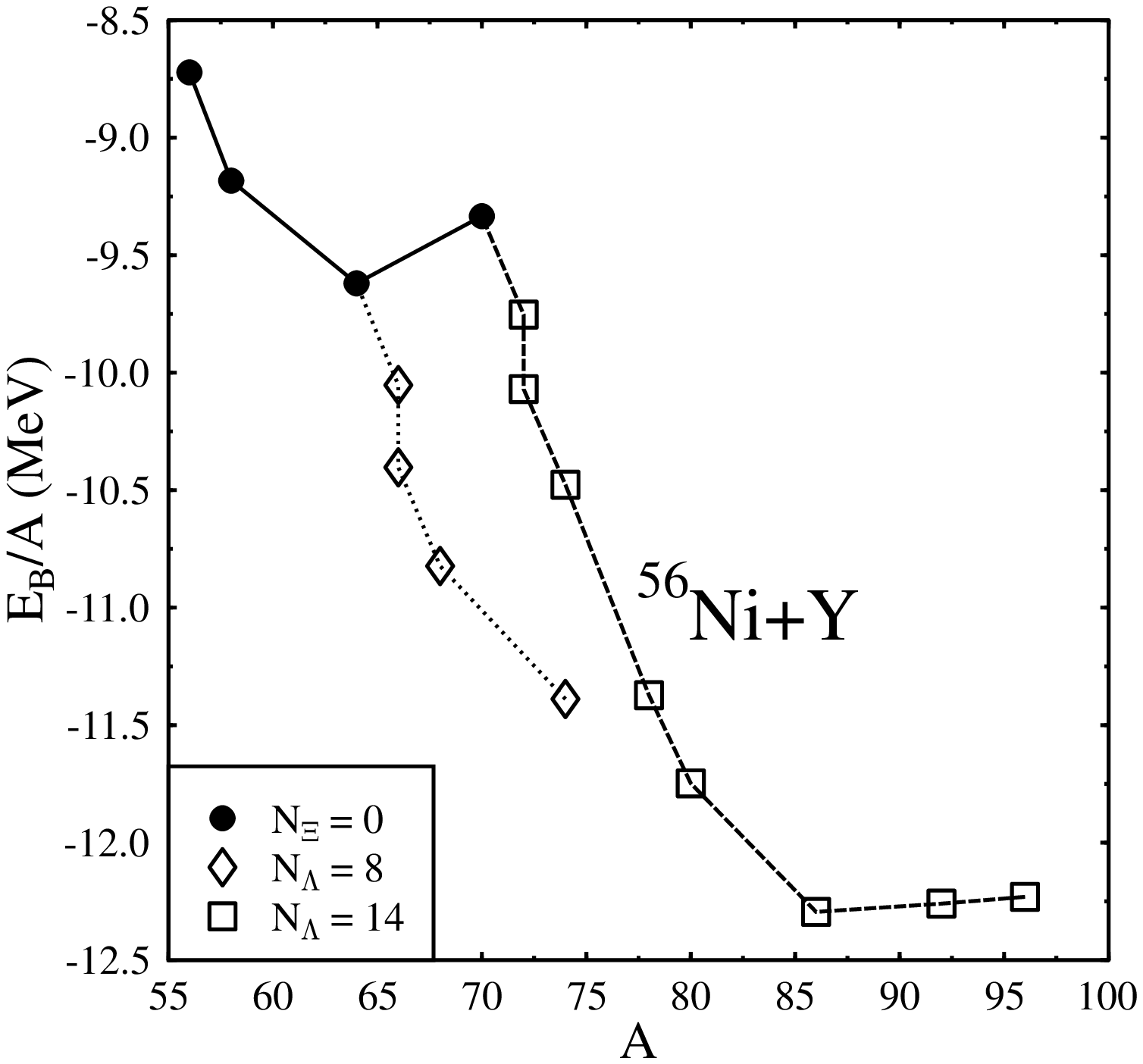}}
\caption{Left: fractions of baryons and leptons in neutron star matter 
calculated in RMF with weak $YY$ potentials \cite{jsb08}. Note that $\Xi^-$ 
hyperons appear at a density range where $\Sigma^-$ hyperons would have 
appeared for an attractive $\Sigma^-$ nuclear potential. Right: binding 
energy of $^{56}$Ni with added $\Lambda$ and $\Xi$ hyperons as a function 
of baryon number $A$ \cite{sbg93}. These particle-stable multistrange states 
decay by weak interactions on a time scale $10^{-10}$ s. } 
\label{fig:sbg} 
\end{figure} 

$\Xi$ hyperons could become stabilized in multi-$\Lambda$ hypernuclei 
once the decay $\Xi N \to \Lambda\Lambda$, which releases $\approx 25$~MeV 
in free space, gets Pauli blocked.{\footnote{With $\approx 80$~MeV release 
in $\Sigma N \to \Lambda N$, however, $\Sigma$ hyperons are unlikely to 
stabilize.}} The onset of $\Xi$ particle-stability would occur 
for $_{\Xi^0\Lambda}^{~~~6}$He or for $_{\Xi^0\Lambda\Lambda}^{~~~~7}$He, 
depending on whether or not $_{\Xi^0}^{~5}{\rm He}$ is bound, and by how much 
(if bound) \cite{sbg94}. Particle stability for $\Xi$ hyperons becomes robust 
with few more $\Lambda$s, even for as shallow $\Xi$-nucleus potentials 
as discussed above. The r.h.s. of Fig.~\ref{fig:sbg} demonstrates that 
$\Xi$s can be added to a core of $^{56}$Ni plus $\Lambda$s, reaching as high 
strangeness fraction as $f_S\equiv -S/A\approx 0.7$ while retaining particle 
stability. This leads to the concept of Strange Hadronic Matter (SHM) 
consisting of equal fractions of protons, neutrons, $\Lambda$, $\Xi^0$ and 
$\Xi^-$ hyperons \cite{sbg93}, with $f_S=1$ as in Strange Quark Matter 
(SQM). Both SHM and SQM provide macroscopic realizations of strangeness, 
but SHM is more plausible phenomenologically, whereas SQM is devoid of any 
experimental datum from which to extrapolate.

\section{${\overline K}$ nuclear interactions and ${\overline K}$ condensation} 
\label{sec:Kbar} 

The $\bar K$-nucleus interaction near threshold comes strongly attractive and 
absorptive in fits to the strong-interaction shifts and widths of $K^-$-atom 
levels~\cite{fg07}, resulting in deep potentials, 
Re~$V^{\bar K}(\rho_0)\sim -(150-200)$ MeV at threshold. 
Chirally based coupled-channel models that fit the low-energy $K^-p$ 
reaction data, and the $\pi\Sigma$ spectral shape of the $\Lambda(1405)$ 
resonance, yield weaker but still very attractive potentials, 
Re~$V^{\bar K}(\rho_0)\sim -100$ MeV, as summarized recently in 
Ref.~\cite{weise08}. A third class, of relatively shallow potentials 
with Re~$V^{\bar K}(\rho_0) \sim -(40-60)$ MeV, was obtained by imposing 
a Watson-like self-consistency requirement \cite{ramoset00}. 

\begin{table} 
\tbl{Calculated $B_{K^-pp}$, mesonic ($\Gamma_{\rm m}$) 
\& nonmesonic ($\Gamma_{\rm nm}$) widths.}
{\begin{tabular}{@{}lccccc@{}}\toprule 
&\multicolumn{2}{c}{$\bar K NN$ single channel} 
&\multicolumn{3}{c}{$\bar K NN - \pi\Sigma N$ coupled channels} \\ 
(MeV) & ATMS~\cite{akaishi02} & AMD~\cite{dote08} & Faddeev~\cite{shevch07} & 
Faddeev~\cite{ikeda07} & variational~\cite{wycech09} \\\colrule 
$B_{K^-pp}$ & $48$ & $17-23$ & $50-70$ & $60-95$ & $40-80$ \\ 
$\Gamma_{\rm m}$ & $61$ & $40-70$  & $90-110$ & $45-80$ & $40-85$ \\ 
$\Gamma_{\rm nm}$ & $12$ & $4-12$ & & & $\sim 20$ \\\botrule 
\end{tabular}} 
\label{tab:kpp} 
\end{table} 

The onset of nuclear (quasi) binding for $K^-$ mesons occurs already with just 
one proton: the $\Lambda(1405)$ which is represented by an $S$-matrix pole 
about 27 MeV below the $K^-p$ threshold. However, in chirally based models, 
the $I=0$ $\bar K N - \pi\Sigma$ coupled channel system exhibits also another 
$S$-matrix pole roughly 12 MeV below threshold and it is this pole that enters 
the effective $\bar KN$ interaction, affecting dominantly the $\bar K$-nucleus 
dynamics \cite{weise08}. The distinction between models that consider 
the twin-pole situation and those that are limited to the $\Lambda(1405)$ 
single-pole framework shows up already in calculations of 
$[\bar K (NN)_{I=1}]_{I=1/2,J^{\pi}=0^-}$, loosely denoted $K^-pp$, which 
is the configuration that maximizes the strongly attractive $I=0~\bar K N$ 
interaction with two nucleons. In Table~\ref{tab:kpp} which summarizes 
$K^-pp$ binding-energy calculations, the $I=0$ $\bar K N$ binding input to 
the ATMS calculation is stronger by about 15 MeV than for the AMD calculation, 
resulting in almost 30 MeV difference. Furthermore, it is clear from 
the `coupled-channel' entries in the table that the explicit use of the 
$\pi\Sigma N$ channel adds about $20 \pm 5$~MeV to the binding energy 
calculated using effective $\bar K N$ potential within a single-channel 
calculation. The experimental state of the art in searching for a $K^-pp$ 
signal was discussed in this Symposium by Yamazaki, and also by Fabbietti 
(FOPI) and Piano (FINUDA). In view of the wide spectrum of predictions in 
Table~\ref{tab:kpp}, new dedicated experiments are welcome; indeed a `day-1' 
experiment at J-PARC on a $^3$He target is scheduled soon \cite{lio08}. 

\begin{figure}[htp] 
\centerline{
\includegraphics[width=5.8cm]{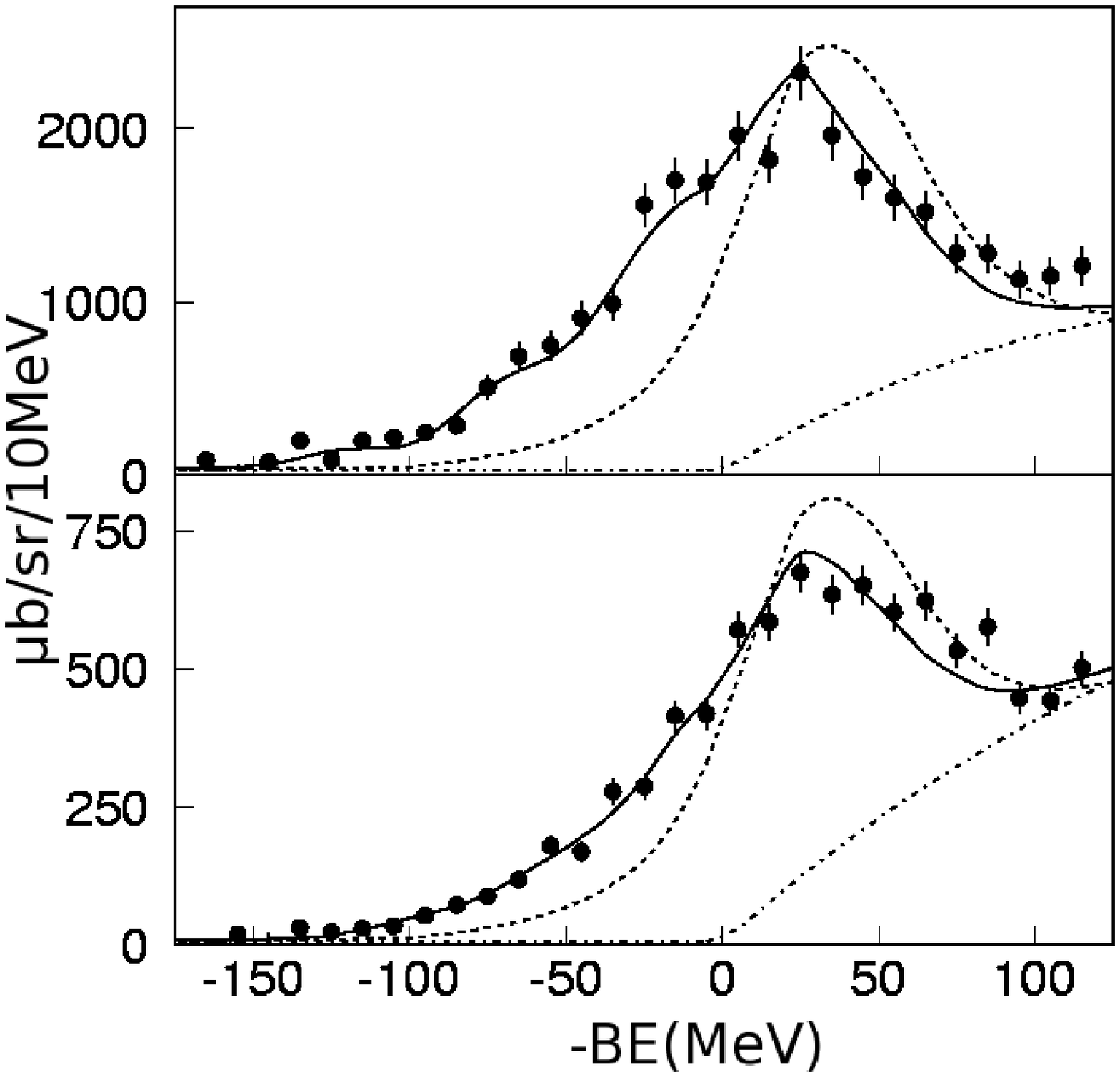} 
\includegraphics[width=5.8cm]{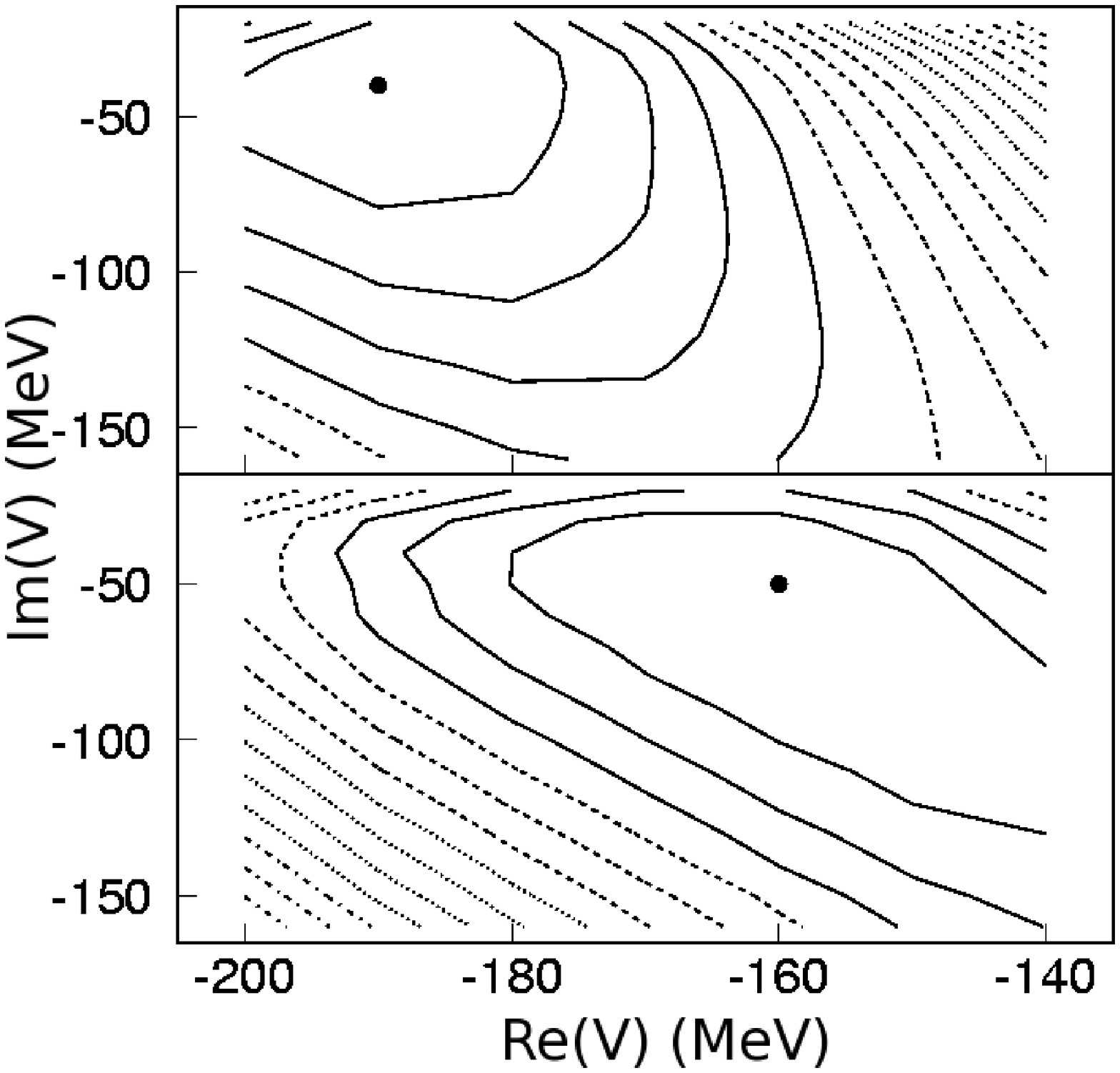}} 
\caption{Missing mass spectra (left) and $\chi^2$ contour plots (right) for 
the inclusive reactions $(K^-,n)$ (upper) and $(K^-,p)$ (lower) at 
$p_{K^-}=1$~GeV/c on $^{12}$C, from Ref.~\cite{kish07}} 
\label{fig:kish} 
\end{figure} 

A fairly new and independent evidence in favor of deep 
$\bar K$-nucleus potentials is provided by $(K^-,n)$ and $(K^-,p)$ 
spectra \cite{kish07} taken at KEK on $^{12}$C, and very recently also 
on $^{16}$O at $p_{K^-}=1$ GeV/c (presented in PANIC08). The $^{12}$C 
spectra are shown in Fig.~\ref{fig:kish}, where the solid lines on the 
left-hand side represent calculations (outlined in Ref.~\cite{yamagata06}) 
using potential depths in the range $160-190$ MeV. The dashed lines 
correspond to using relatively shallow potentials of depth about 60 MeV 
which I consider therefore excluded by these data. Although the potentials 
that fit these data are sufficiently deep to support strongly-bound antikaon 
states, a fairly sizable extrapolation is required to argue for 
$\bar K$-nuclear quasibound states at energies of order 100 MeV below 
threshold, using a potential determined largely near threshold. Furthermore, 
the best-fit Im~$V^{\bar K}$ depths of $40-50$ MeV imply that $\bar K$-nuclear 
quasibound states are broad, as studied in Refs.~\cite{mares06,gazda07}. 

\begin{figure}[htp]
\centerline{
\includegraphics[width=6.0cm,height=5.5cm]{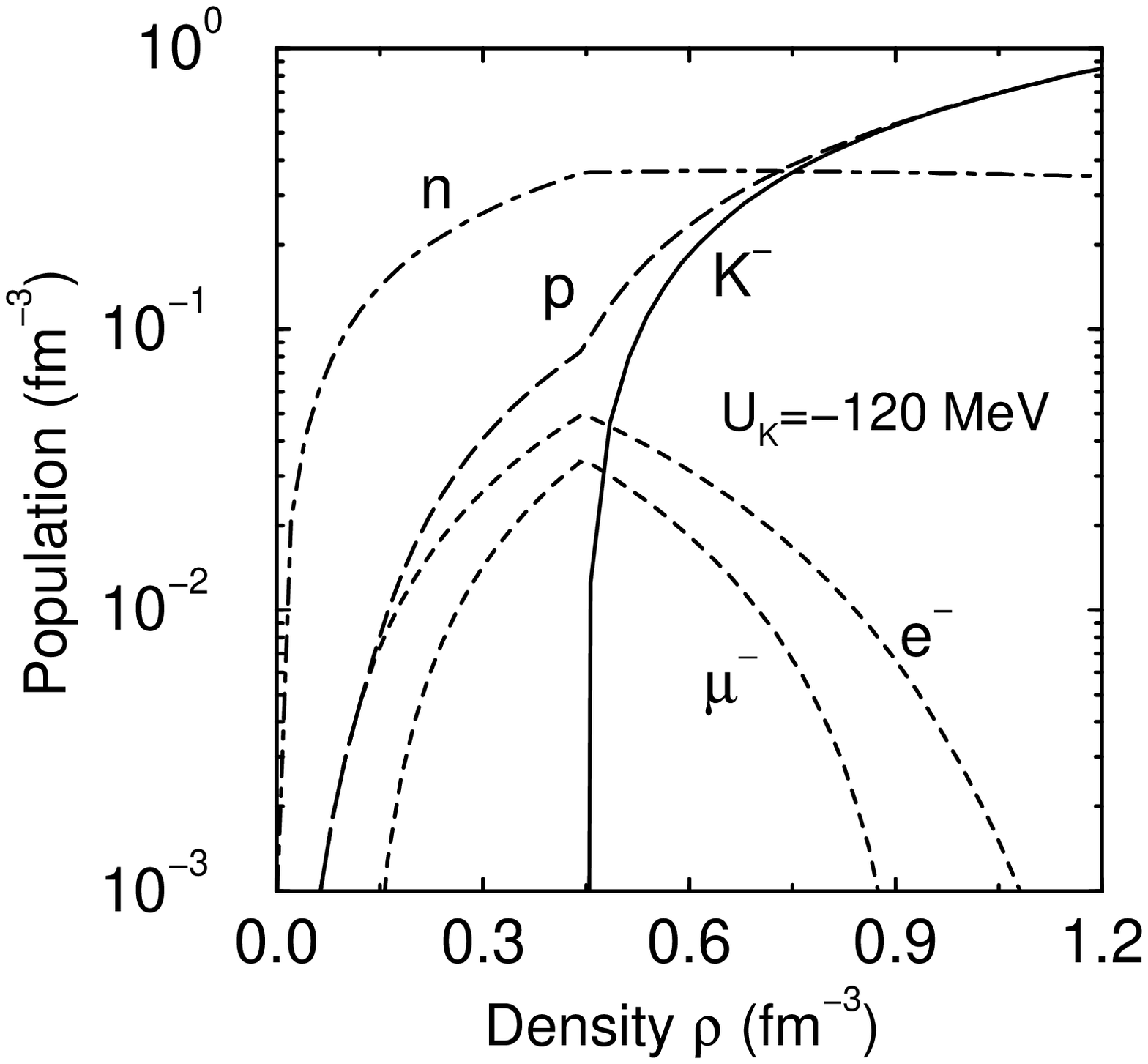} 
\includegraphics[width=5.0cm,height=5.0cm]{multik40ca.eps}} 
\caption{Left: calculated neutron-star population as a function of nucleon 
density, from Ref.~\cite{glend99} The neutron density stays nearly constant 
once kaons condense. 
Right: calculated separation energies $B_{K^-}$ in multi-$K^-$ nuclei based 
on $^{40}{\rm Ca}$ as a function of the number $\kappa$ of $K^-$ mesons 
in several nuclear RMF models with two choices of parameters fixed for 
$\kappa=1$, from Ref.~\cite{gazda08} See also Mare\v{s}' talk in this 
Symposium.}
\label{fig:jsb} 
\end{figure} 

A robust consequence of the sizable $\bar K$-nucleus attraction is that $K^-$ 
condensation occurs in neutron stars at about 3 times nuclear matter density, 
as shown on the l.h.s. of Fig.~\ref{fig:jsb}. Comparing it with the l.h.s. of 
Fig.~\ref{fig:sbg}, also for neutron stars, but where strangeness materialized 
through hyperons, one may ask whether or not the r.h.s of Fig.~\ref{fig:sbg}, 
for finite nuclei, also offers an analogy: do $\bar K$ mesons condense 
in nuclear matter? This question was posed and answered, negatively, 
in Ref.~\cite{gazda08} calculating multi-$\bar K$ nuclear configurations. 
The r.h.s. of Fig.~\ref{fig:jsb} demonstrates a remarkable saturation of 
$K^-$ separation energies $B_{K^-}$ calculated in multi-$K^-$ nuclei 
$^{40}{\rm Ca}+\kappa K^-$, independently of the applied RMF model. 
The saturation values of $B_{K^-}$ do not allow conversion of hyperons 
to $\bar K$ mesons through the strong decays $\Lambda \to p + K^-$ 
or $\Xi^- \to \Lambda + K^-$ in multi-strange hypernuclei, which therefore 
remain the lowest-energy configuration for multi-strange systems. This 
provides a powerful argument against $\bar K$ condensation in the laboratory, 
under strong-interaction equilibrium conditions \cite{gazda08}. It does not 
apply to kaon condensation in neutron stars, where equilibrium configurations 
are determined by weak-interaction conditions.

\section*{Acknowledgments} 
Thanks are due to John Millener for instructive correspondence on the analysis 
of the $\gamma$-ray experiments, and to the Organizers of SENDAI08 for their 
gracious hospitality.

\end{document}